\documentclass[aps,prc,twocolumn,groupedaddress,showpacs,10pt,dvipsnames]{revtex4-1}

\usepackage[utf8]{inputenc}
\usepackage{amsfonts}
\usepackage{amsmath}
\usepackage{amssymb}
\usepackage{bm}
\usepackage[pdftex]{graphicx}
\usepackage{xcolor}
\usepackage[normalem]{ulem}

\begin{document}

\newcommand{\tj}[6]{ \begin{pmatrix}
  #1 & #2 & #3 \\
  #4 & #5 & #6
 \end{pmatrix}}


\title{Two-nucleon emitters within a pseudostate method: The case of $^6$Be and $^{16}$Be.} 




\author{J. Casal}
\email{casal@ectstar.eu}
\affiliation{European Centre for Theoretical Studies in Nuclear Physics and Related Areas (ECT$^*$), Villa Tambosi, Strada delle Tabarelle 286, I-38123 Villazzano (Trento), Italy}


\date{\today}

\begin{abstract}
 \begin{description}
  \item[Background] Since the first experimental observation, two-nucleon radioactivity has gained renewed attention over the past fifteen years. The $^6$Be system is the lightest two-proton ground-state emitter, while $^{16}$Be was recently proposed to be the first two-neutron ground-state emitter ever observed. A proper understanding of their properties and decay modes requires a reasonable description of the three-body continuum.
  \item[Purpose] Study the ground-state properties of $^6$Be and $^{16}$Be within a general three-body model and investigate their nucleon-nucleon correlations in the continuum.
  \item[Method] The pseudostate (PS) method in hyperspherical coordinates, using the analytical transformed harmonic oscillator (THO) basis for three-body systems, is used to construct the $^6$Be and $^{16}$Be ground-state wave functions. These resonances are approximated as a stable PS around the known two-nucleon separation energy. Effective core-$N$ potentials, constrained by the available experimental information on the binary subsystems $^5$Li and $^{15}$Be, are employed in the calculations. 
  \item[Results] The ground state of $^{16}$Be is found to present a strong dineutron configuration, with the valence neutrons occupying mostly an $l=2$ state relative to the core. The results are consistent with previous $R$-matrix calculations for the actual continuum. The case of $^6$Be shows a clear symmetry with respect to its mirror partner, the two-neutron halo $^6$He: The diproton configuration is dominant, and the valence protons occupy an $l=1$ orbit.
  \item[Conclusions] The PS method is found to be a suitable tool in describing the properties of unbound $\text{core}+N+N$ ground states. For both $^{16}$Be and $^6$Be, the results are consistent with previous theoretical studies and confirm the dominant dinucleon configuration. This favors the picture of a correlated two-nucleon emission.
 \end{description} 
\end{abstract}


\maketitle


\section{Introduction}
Exotic nuclei far from stability give rise to unusual properties and decay modes~\cite{pfutzner12}. In the past few decades, the advances in radioactive beam physics has enabled the study and characterization of nuclear systems close to the neutron and proton driplines. 
Large efforts have been devoted to understanding the properties of two-neutron halo nuclei~\cite{Zhukov93,Tanihata13}. These are Borromean systems, in which all binary subsystems do not form bound states. Theoretical investigations within $\text{core}+n+n$ models indicate that the correlations between the valence neutrons play a fundamental role in shaping the properties of two-neutron halo nuclei~\cite{Zhukov93,Nielsen01,kikuchi16}. 

The evolution of these correlations beyond the driplines has implications for two-nucleon radioactivity. First proposed for two-proton decays in the sixties~\cite{Goldansky60}, this topic gained renewed attention after the first experimental observation of the correlated emission from the ground state of $^{45}$Fe~\cite{giovinazzo02,pfutzner02}. Since then, other examples of two-proton emitters have been confirmed, e.g. $^{54}$Zn~\cite{blank05}, $^{19}$Mg~\cite{muhka07} or $^6$Be~\cite{grigorenko09}. More recently, the case of two-neutron emission has also been observed from $^{16}$Be~\cite{spyrou12}, $^{26}$O~\cite{kohley13} and $^{24}$O~\cite{jones15}. 

The decay paths for two-nucleon emitters can follow different mechanisms. If there is a narrow state available in the intermediate nucleus, i.e., below the ground state of the parent system, the process is expected to proceed sequentially. On the contrary, if this sequential decay is not energetically possible and the parent nucleus present a strong correlation between the two nucleons prior to emission, a simultaneous ``dinucleon'' decay takes place (see Ref.~\cite{lovell17} and references therein). When these extreme pictures do not apply, the concept of true three-body ``democratic''~\cite{pfutzner12} decay is introduced. The boundaries between these two- and three-body dynamics are still not clear, specially in the decay of excited states, although there have been recent developments~\cite{egorova12}. In this context, three-body models are a natural choice to study the nucleon-nucleon correlation and decay modes.

Very exotic beryllium isotopes offer a good opportunity to study two-nucleon correlations. On the proton-rich side, $^6$Be is known to be the lightest two-proton emitter in its original sense~\cite{Goldansky60}: The intermediate $^5$Li states are not accessible for sequential decay from the ground state of $^6$Be~\cite{grigorenko09,egorova12}. On the neutron-rich side, the case of $^{16}$Be was claimed to be the first experimental observation of a ground-state decay showing a clear signature of correlated dineutron emission~\cite{spyrou12}. Three-body models in terms of $^4\text{He}+p+p$ and $^{14}\text{Be}+n+n$ have been recently used to analyze the structure of these unbound systems~\cite{grigorenko09,egorova12,lovell17,oishi17}. This requires a proper description of $\text{core}+N+N$ continuum states. The three-body continuum problem for systems comprising a single charged particle can be solved, for instance, using the hyperspherical $R$-matrix theory~\cite{lovell17}. The extension for systems involving the Coulomb interaction is not an easy task, as the asymptotic behavior for these systems is not known in general. To deal with this problem, very involved procedures are needed~\cite{alvarezrodriguez08,Nguyen12,Ishikawa13}, not free from uncertainties. 

\newpage
An alternative is the so-called pseudostate (PS) method~\cite{SiegertPS}, which consists in diagonalizing the Hamiltonian in a complete set of square-integrable functions. This provides the bound states of the system, and also a discrete representation of the continuum. In this context, a variety of bases have been proposed for two-body~\cite{HaziTaylor70,Matsumoto03,MRoGa04,AMoro09} and three-body systems~\cite{Desc03,Matsumoto04,MRoGa05,JCasal13}. Lately, the PS method in hyperspherical coordinates~\cite{Zhukov93,Nielsen01} has been successfully applied to describe the structure properties and reaction dynamics of three-body nuclei (e.g.~\cite{JCasal14,Descouvemont15,JCasal15,JCasal16,casalplb17,gomezramosplb17}). This approach, involving a standard eigenvalue problem, is computationally simpler than the calculation of actual continuum states.

It is the purpose of this work to study the ground-state properties of $^{16}$Be and $^6$Be by means of the PS method in a three-body ($\text{core}+N+N$) scheme. Calculations are constrained by the experimental information on the binary subsystems $^{15}$Be ($^{14}\text{Be}+n$)~\cite{snyder13} and $^5$Li ($^{4}\text{He}+p$)~\cite{tilley02}, and the known two-nucleon separation energies in $^{16}$Be~\cite{spyrou12}, $^6$Be~\cite{tilley02}. The validity of the discretization is assessed by comparing with previous theoretical studies, and the results are analyzed in terms of two-nucleon correlations.

The paper is structured as follows. In Sec.~\ref{sec:formalism}, the three-body formalism used in this work is presented. Results for $^{16}$Be and $^6$Be are shown in Sec.~\ref{sec:application}, where the reliability of the theoretical approach is discussed. Finally, Sec.~\ref{sec:conclusions} summarizes the main conclusions and outlines possible further applications.

\section{Hyperspherical Harmonics (HH) Formalism}
\label{sec:formalism}

\begin{figure}
\centering
 \includegraphics[width=\linewidth]{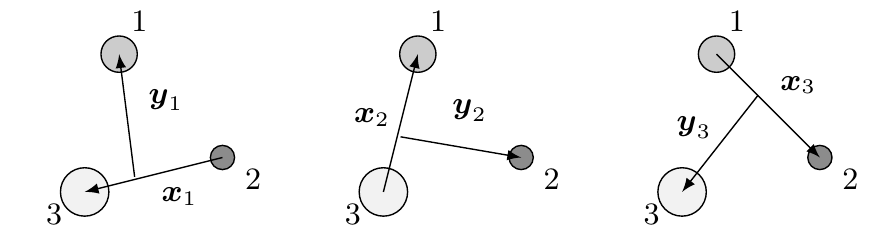}
 \caption{The three sets of scaled Jacobi coordinates.}
 \label{fig:sets}
\end{figure}

Three-body systems can be described using Jacobi coordinates $\{\boldsymbol{x}_k,\boldsymbol{y}_k\}$, where the label $k=1,2$ or 3 indicates one of the three coordinate choices in Fig.~\ref{fig:sets}. In these sets, the variable $\boldsymbol{x}_k$ is proportional to the relative coordinate between two particles and $\boldsymbol{y}_k$ is proportional to the distance from the center of mass of the $x$-subsystem to the third particle. The scaling factors between physical distances and Jacobi coordinates are given by~\cite{Zhukov93}
\begin{equation}
\boldsymbol{x}_k=\boldsymbol{r}_{x_k}\sqrt{\frac{A_iA_j}{A_i+A_j}},
\end{equation}
and
\begin{equation}
\boldsymbol{y}_k=\boldsymbol{r}_{y_k}\sqrt{\frac{A_k(A_i+A_j)}{A}},
\end{equation}
where $A=A_i+A_j+A_k$ is the total mass number and $\{i,j,k\}$ are in a cyclic order. It is then clear that the Jacobi-$k$ set corresponds to the system where particles $(i,j)$ are related by the $x$-coordinate. 
From Jacobi coordinates, the hyperspherical coordinates  $\{\rho,\alpha_k,\widehat{x}_k,\widehat{y}_k\}$ can be introduced. Here, the hyper-radius $(\rho)$ and the hyperangle $(\alpha_k)$ are given by
\begin{align}
\rho = & \sqrt{x_k^2 + y_k^2}, \\
\alpha_k = & \tan\left(\frac{x_k}{y_k}\right),
\label{eq:jachyp}
\end{align}
and $\{\widehat{x}_k,\widehat{y}_k\}$ are the two-dimensional angular variables related to $\{\boldsymbol{x}_k,\boldsymbol{y}_k\}$. Note that, while the hyperangle depends on $k$, the hyper-radius does not.

In the hyperspherical harmonic (HH) formalism, the eigenstates of the system are expanded in hyperspherical coordinates as 
\begin{equation}
 \Psi^{j\mu}(\rho,\Omega)=\frac{1}{\rho^{5/2}}\sum_{\beta} \chi_{\beta}^{j}(\rho)\mathcal{Y}_{\beta}^{j\mu}(\Omega),
 \label{eq:wf}
\end{equation}
where $\Omega\equiv\{\alpha,\widehat{x},\widehat{y}\}$ is introduced for the angular dependence. For simplicity, the label $k$ has been omitted, assuming a fixed Jacobi set. Here, $\beta\equiv\{K,l_x,l_y,l,S_x,j_{ab}\}$ is a set of quantum numbers referred to as channel, where $K$ is the hypermomentum, $l_x$ and $l_y$ are the orbital angular momenta associated with the Jacobi coordinates $\boldsymbol{x}$ and $\boldsymbol{y}$, respectively, $l$  is the total orbital angular momentum ($\boldsymbol{l}=\boldsymbol{l_x}+\boldsymbol{l_y}$), $S_x$ is the spin of the particles related by the coordinate $\boldsymbol{x}$, and $j_{ab}$ results from the coupling $\boldsymbol{j_{ab}}=\boldsymbol{l}+\boldsymbol{S_x}$. By denoting by $I$  the spin of the third particle, which is assumed to be fixed, the total angular momentum is $\boldsymbol{j}=\boldsymbol{j_{ab}} + \boldsymbol{I}$. The angular functions in Eq.~(\ref{eq:wf}), $\mathcal{Y}_{\beta}^{j\mu}(\Omega)$, are states of good total angular momentum, which are expanded as~\cite{Zhukov93}
\begin{equation}
\mathcal{Y}_{\beta}^{j\mu}(\Omega)=\left\{\left[\Upsilon_{Klm_l}^{l_xl_y}(\Omega)\otimes\kappa_{s_x}\right]_{j_{ab}}\otimes\phi_I\right\}_{j\mu},
\label{eq:Upsilon}
\end{equation}
where $\Upsilon_{Klm_l}^{l_xl_y}$ are the hyperspherical harmonics. These are the analytical eigenfunctions of the hypermomentum operator $\widehat{K}$, given by
\begin{equation}
\Upsilon_{Klm_l}^{l_xl_y}(\Omega)=\varphi_K^{l_xl_y}(\alpha)\left[Y_{l_x}(\boldsymbol{x})\otimes Y_{l_y}(\boldsymbol{y})\right]_{lm_l},
\label{eq:HH}
\end{equation}
\begin{equation}
\begin{split}
\varphi_K^{l_xl_y}(\alpha) & = N_{K}^{l_xl_y}\left(\sin\alpha\right)^{l_x}\left(\cos\alpha\right)^{l_y}\\&\times P_n^{l_x+\frac{1}{2},l_y+\frac{1}{2}}\left(\cos 2\alpha\right),
\label{eq:varphi}
\end{split}
\end{equation}
with $P_n^{a,b}$ a Jacobi polynomial of order $n=(K-l_x-l_y)/2$ and $N_K^{l_xl_y}$ a normalization constant. With the above definition, the three-body Schrödinger equation leads to a set of coupled hyperradial equations
\begin{equation}
\begin{split}
&\left[-\frac{\hbar^2}{2m}\left(\frac{d^2}{d\rho^2}-\frac{15/4+K(K+4)}{\rho^2}\right)-\varepsilon\right]\chi_{\beta}^{j}(\rho)\\&+\sum_{\beta'}V_{\beta'\beta}^{j\mu}(\rho) \chi_{\beta'}^{j}(\rho)=0,
\end{split}
\label{eq:coupled}
\end{equation}
where $V_{\beta'\beta}^{j\mu}(\rho)$ are the coupling potentials defined as
\begin{equation}
V_{\beta'\beta}^{j\mu}(\rho)=\left\langle \mathcal{Y}_{\beta }^{j\mu}(\Omega)\Big|V_{12}+V_{13}+V_{23} \Big|\mathcal{Y}_{\beta'}^{ j\mu}(\Omega) \right\rangle.
\end{equation}

In this work, the system given by Eq.~(\ref{eq:coupled}) is replaced by a standard eigenvalue problem by using the pseudo-state (PS) method~\cite{SiegertPS}, Here, as in Refs.~\cite{JCasal13,JCasal14,JCasal15,JCasal16},  the analytical transformed harmonic oscillator (THO) basis is used. The radial functions are expanded as
\begin{equation}
 \chi_{\beta}^{j}(\rho) = \sum_{i} C_{i\beta}^{j} U_{i\beta}^\text{THO}(\rho),
 \label{eq:expandTHO}
\end{equation}
where $i$ denotes the hyperradial excitation and $C_{i\beta}^{j}$ are just the diagonalization coefficients. Therefore, the wave functions~(\ref{eq:wf}) involve infinite sums over $\beta$ and $i$. However, calculations are typically truncated at maximum hypermomentum $K_{max}$ and $i_{max}$ hyperradial excitations in each channel. These parameters have to be large enough to provide converged results. 

The THO basis functions in Eq.~(\ref{eq:expandTHO}) are obtained from the harmonic oscillator (HO) functions using a local scale transformation, $s(\rho)$, satisfying the relationship
\begin{equation}
  U_{i\beta}^{\text{THO}}(\rho)=\sqrt{\frac{ds}{d\rho}}U_{iK}^{\text{HO}}[s(\rho)].
\label{eq:R}
\end{equation}
This transformation keeps the simplicity of the HO functions, but converts their Gaussian asymptotic behavior into an exponential one. This provides a suitable representation of bound and resonant states to calculate structure and scattering observables. For this purpose. the analytical form proposed by Karataglidis \textsl{et al.}~\cite{Karataglidis} can be used,
\begin{equation}
s(\rho) = \frac{1}{\sqrt{2}b}\left[\frac{1}{\left(\frac{1}{\rho}\right)^{4} +
\left(\frac{1}{\gamma\sqrt{\rho}}\right)^4}\right]^{\frac{1}{4}}.
\label{eq:LST}
\end{equation}
Note that the THO hyperradial wave functions depend, in general, on all the quantum numbers included in a channel $\beta$, although the HO hyperradial wave functions only depend on the hypermomentum $K$. The preceding transformation depends on parameters $\gamma$ and $b$. The most interesting feature of the analytical THO method is that the ratio $\gamma/b$ governs the asymptotic behavior of the basis functions and controls the density of PSs as a function of the energy. This allows us to select an optimal basis depending on the system or observable under study~\cite{JCasal13}. In order to study the properties of a single three-body resonance, the Hamiltonian can be diagonalized using a THO basis with a small hyperradial extension. This gives a representation of the continuum characterized by a low level density, so that the resonant behavior can be associated with a single PS. 
Examples of this approach have been previously reported, for instance, to study the properties of the 2$^+$ resonance in $^6$He~\cite{MRoGa05,JCasal13} or the 5/2$^-$ resonance in $^9$Be~\cite{Descouvemont15}. Here, the spatial distribution of the valence nucleons in unbound $\text{core}+N+N$ states is analyzed in terms of the ground-state probability written in the Jacobi-1 set,
\begin{equation}
P(x,y)=x^2y^2\int \big|\Psi^{j\mu}(\boldsymbol{x},\boldsymbol{y})\big|^2 d\widehat{x}d\widehat{y},
\label{eq:Pxy}
\end{equation}
where the wave function has been transformed back to Jacobi coordinates, and, after scaling to the relative distances $r_x\equiv r_{N\text{-}N}$ and $r_y\equiv r_{\text{core-}(NN)}$, it satisfies the normalization relationship
\begin{equation}
\int P(r_x,r_y)dr_xdr_y=1.
\label{eq:Prxry}
\end{equation}

\section{Application to exotic beryllium isotopes}
\label{sec:application}

\begin{figure}[t]
	\centering
	\includegraphics[width=0.8\linewidth]{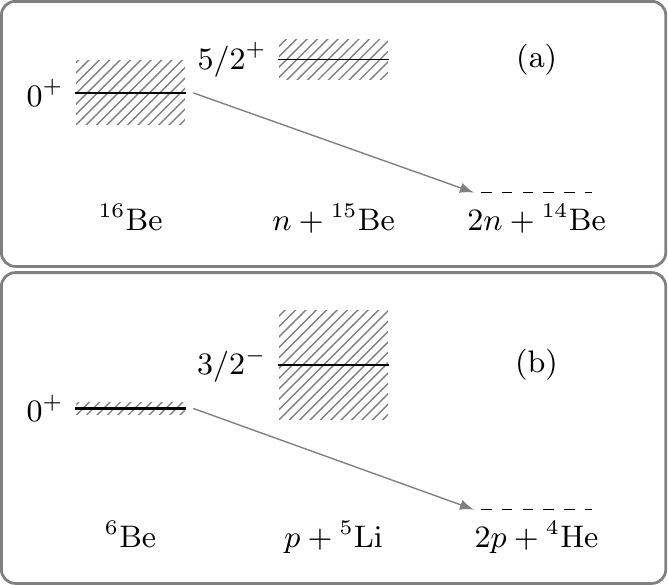} 
	\caption{Two-nucleon decay paths for (a) $^{16}$Be and (b) $^6$Be.}
	\label{fig:paths}
\end{figure}

The only stable beryllium isotope, $^9$Be, is already a weakly bound system~\cite{tilley04}. Exotic $Z=4$ isotopes form bound states from $^7$Be to $^{14}$Be (with the exception of the unbound systems $^8$Be and $^{13}$Be). Beyond the driplines, $^{16}$Be and $^6$Be ground states have been observed as $0^+$ resonances characterized by two-nucleon separation energies  $S_{2n}(^{16}\text{Be})=-1.35(10)$ MeV~\cite{spyrou12} and $S_{2p}(^6\text{Be})=-1.372(5)$ MeV~\cite{tilley02}. Their widths are 0.8 and 0.092 MeV, respectively, although a much narrower state, 0.17 MeV, was found for $^{16}$Be in recent calculations~\cite{lovell17}. The discrepancy was attributed to the effect of the experimental resolution. The properties of the relevant binary subsystems $^{15}$Be and $^5$Li have also been measured. The ground state of $^{15}$Be is a $d_{5/2}$ state 1.8(1) MeV above the neutron separation threshold and has a width of 0.58(20) MeV~\cite{snyder13}. On the other hand, the $p_{3/2}$ state in $^5$Li is unbound with respect to the proton emission by 1.96(5) MeV, and its accepted width is 1.5 MeV~\cite{tilley02}. Therefore, the sequential two-nucleon emission from the ground state of $^{16}$Be ($^6$Be) is (mostly) unaccessible, as shown in Fig.~\ref{fig:paths}. This favors a simultaneous two-nucleon emission, either in the form of a ``dinucleon'' or in a true three-body (democratic) decay~\cite{pfutzner12}. 

\begin{table}[b]
	\centering
	\begin{tabular}{ccccc}
		\toprule
		& & $n$-$^{14}$Be~\cite{lovell17} & & $p$-$^{4}$He \\
		\colrule
		$V_c(l=0)$ & & $-$26.18 & & $+$48.00 \\
		$V_c(l=1)$ & & $-$30.50 & & $-$42.35 \\
		$V_c(l=2)$ & & $-$42.73 & & $-$21.50 \\
		$V_{\rm so}(l=1)$ & & $-$10.00 & & $-$40.00\\
		$V_{\rm so}(l=2)$ & & $-$33.77 & & $-$40.00\\
		$a_c$ & & 0.65 & & 0.70\\
		$a_{so}$ & & 0.65 & & 0.35\\
		$R_c$ & & 3.02 & & 2.00\\
		$R_{\rm so}$ & & 3.02 & & 1.50\\	
		$R_{\rm Coul}$ & & $-$ & & 2.00\\			
		\botrule
	\end{tabular}
	\caption{Parameters of the binary potentials for $^{15}$Be and $^5$Li corresponding to Woods-Saxon geometries. The spin-orbit terms follow the typical derivative form using the definition from Ref.~\cite{IJThompson04}. The potential depth ($V$) is given in MeV, while the radius ($R$) and diffuseness ($a$) are provided in fm. $R_{\rm Coul}$ is the hard-sphere Coulomb radius.
	}
	\label{tab:pot}
\end{table}

Three-body $\text{core}+N+N$ descriptions require, as input, a nucleon-nucleon interaction and realistic $\text{core}+N$ potentials. For the former, 
in this work the GPT nucleon-nucleon potential~\cite{GPT} is employed, including central, spin-orbit and tensor terms. This potential, although simpler than the robust Reid93~\cite{reid93} or AV18~\cite{av18} interactions, reproduces $NN$ observables up to 300 MeV. This makes it suitable for three-body calculations~\cite{MRoGa05,JCasal13,lovell17}. The $^{14}\text{Be}+n$ and $^4\text{He}+p$ potentials are adjusted to reproduce the position of the $^{15}$Be and $^5$Li ground states, respectively. These are $l$-dependent Woods-Saxon potentials with central and spin-orbit terms, whose parameters are given in Table~\ref{tab:pot}. Note that, in this work, the $^{14}\text{Be}+n$ interaction is the same used in Ref.~\cite{lovell17}, while the $^{4}\text{He}+p$ potential is essentially the one used in Refs.~\cite{MRoGa05,JCasal13} for the $^{4}\text{He}+n$ case but including also the Coulomb repulsion. The later is a shallow potential, in the sense that the $1s_{1/2}$ Pauli state has been removed by introducing a repulsive $l=0$ term. However, the $^{14}\text{Be}+n$ potential gives rise to $1s_{1/2}$, $1p_{3/2}$ and $1p_{1/2}$ bound states which represent the neutron-occupied orbitals of the core. These states have to be projected out for the $^{14}\text{Be}+n+n$ three-body calculations, and this can be achieved, as in Ref.~\cite{lovell17}, by using a supersymmetric transformation~\cite{Baye87}. Note that the treatment of the Pauli principle in three-body systems is a delicate issue, and the inert core approximation with the 1$s$ and 1$p$ states permanently occupied might not be a very realistic model to describe $^{15,16}$Be. Effects not explicitly included within this strict three-body model could affect the properties of $^{16}$Be.

\begin{figure}[t]
	\centering
	\includegraphics[width=\linewidth]{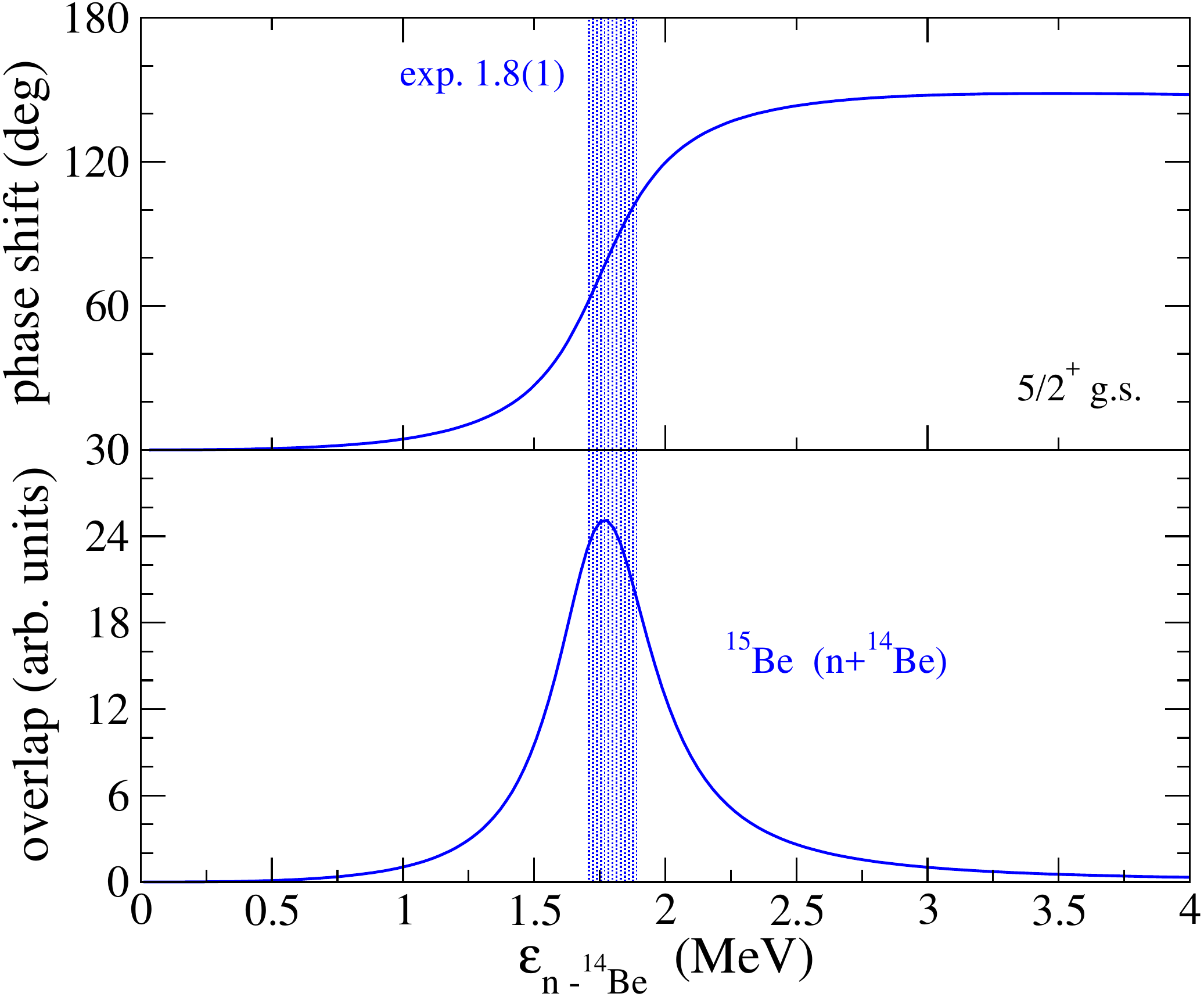} 
	\caption{Upper panel: $n$-$^{14}$Be phase shifts for 5/2$^+$ states. Lower panel: Overlaps between $^{15}$Be continuum states and the $^{16}$Be ground state, with the maximum representing the resonance position. The shaded area corresponds to the experimental value~\cite{snyder13}.}
	\label{fig:15Be}
\end{figure}

\begin{figure}[t]
	\centering
	\includegraphics[width=\linewidth]{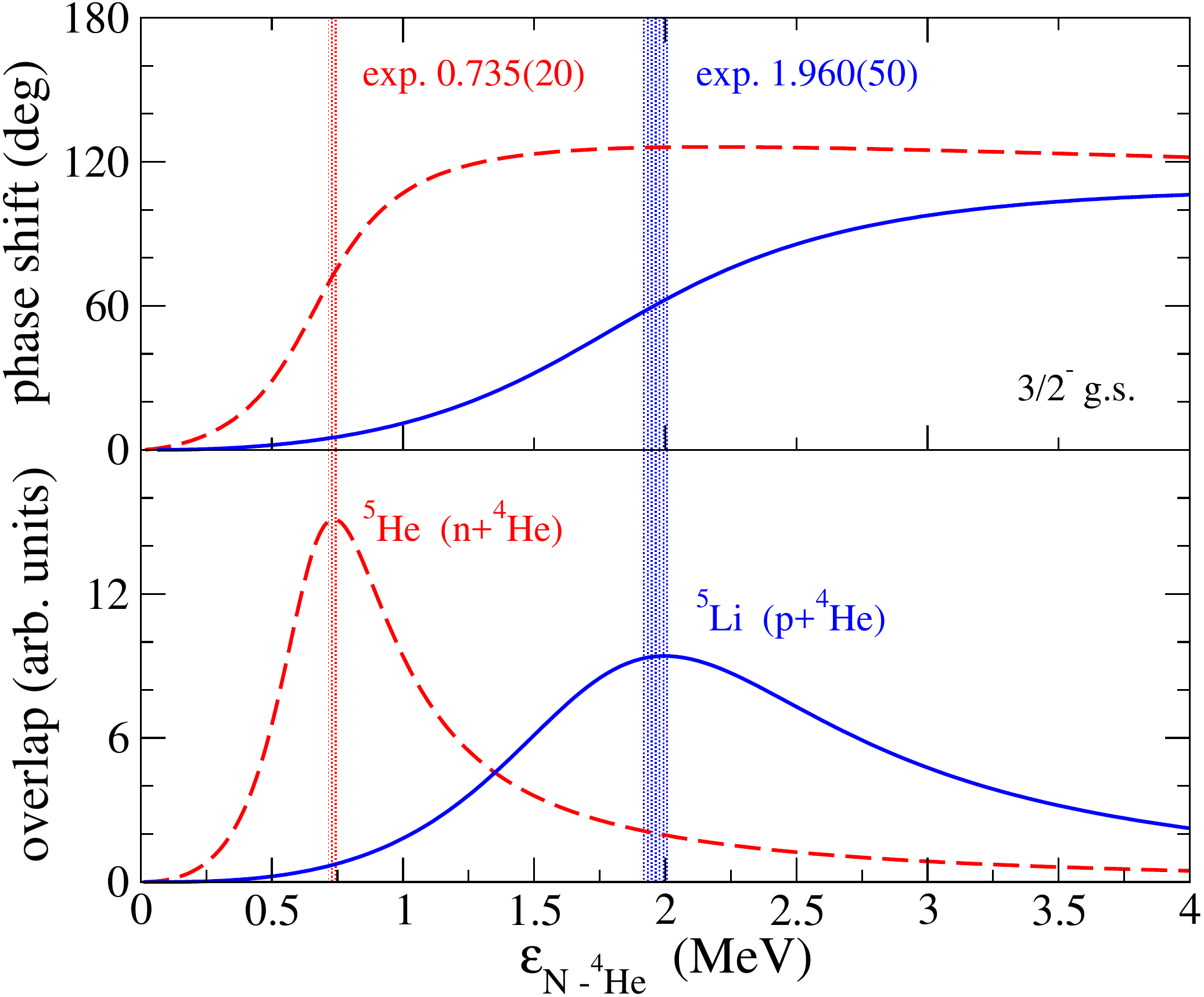} 
	\caption{The same as Fig.~\ref{fig:15Be} but corresponding to the $p$-$^{4}$He (solid) and $n$-$^{4}$He (dashed) 3/2$^-$ continuum, together with the experimental data on $^{5}$Li and $^{5}$He ground-state energies~\cite{tilley02}.}
	\label{fig:5Li}
\end{figure}

The $N$-core phase shifts corresponding to the potentials given in Table~\ref{tab:pot} are shown in the upper panels of Figs.~\ref{fig:15Be} and ~\ref{fig:5Li} for $^{15}$Be(5/2$^+$) and $^5$Li(3/2$^-$) states, respectively. In the lower panels, the position of the two-body resonances can be associated with the maximum of the overlaps between $^{15}$Be ($^5$Li) continuum states and the three-body ground state of $^{16}$Be ($^6$Be). Details about these three-body calculations are given in the following sections. In these figures, vertical lines represent the experimental position of the resonances to which the interactions have been adjusted. For completeness, in Fig.~\ref{fig:5Li}, the phase shifts for $n$-$^4$He scattering as well as the corresponding overlaps are shown together with those for $p$-$^4$He. These have been obtained by just switching off the Coulomb interaction in the binary potential. It is clear that both systems, $^5$He and $^5$Li, can be described using the same $N$-core potential, and this enables the description of the mirror nuclei $^6$Be and $^6$He using the same three-body Hamiltonian except for the Coulomb part. Details are presented in section~\ref{sec:application2}.

\subsection{$2n$ configuration in $^{16}$Be}
\label{sec:application1}

The 0$^+$ states in $^{16}$Be ($^{14}\text{Be}+n+n$) are computed in the Jacobi-1 set, where the two valence neutrons outside a $^{14}$Be core are related by the $\boldsymbol{x}$ coordinate. Since three-body models are an approximation to the full many-body picture, realistic binary interactions alone are typically insufficient to reproduce the known spectra~\cite{MRoGa05,RdDiego10,IJThompson04,JCasal14}. It is then customary to include also a simple hyperradial three-body force, whose parameters can be fixed to reproduce the (known) three-body energies without distorting the structure of the states. In this work, as in Ref.~\cite{JCasal16}, a Gaussian form is adopted,
\begin{equation}
V_{\rm 3b}(\rho)=v_{\rm 3b} \exp(-\rho/\rho_{\rm 3b})^2.
\label{eq:3b}
\end{equation}
Using $\rho_{\rm 3b}=6$ fm and $v_{\rm 3b}=-1$ MeV, a low-lying resonance around the two-neutron separation energy of $^{16}$Be is obtained. Note that this three-body force, with different geometry and parameters, was also employed in the previous $^{16}$Be three-body calculation of Ref.~\cite{lovell17}. 

\begin{figure}
	\centering
	\includegraphics[width=0.9\linewidth]{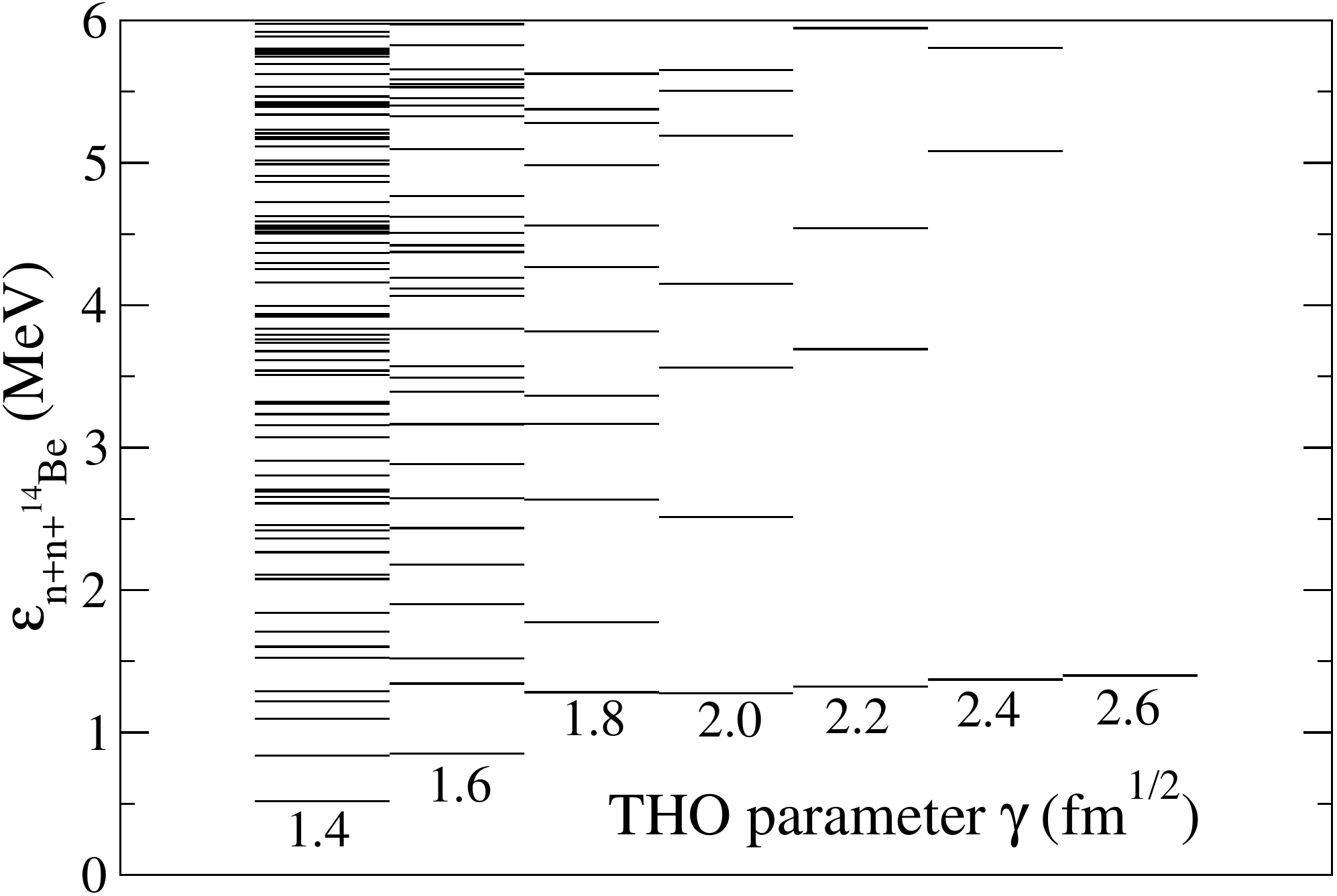} 
	\caption{$^{16}$Be spectra as a function of the THO parameter $\gamma$, which controls the level density after diagonalization. Calculations are provided for $b=0.7$ fm, $K_{max}=30$ and $i_{max}=15$. A stable PS around 1.3 MeV can be clearly identified.}
	\label{fig:stabilization}
\end{figure}

\begin{figure}
	\centering
	\includegraphics[width=0.9\linewidth]{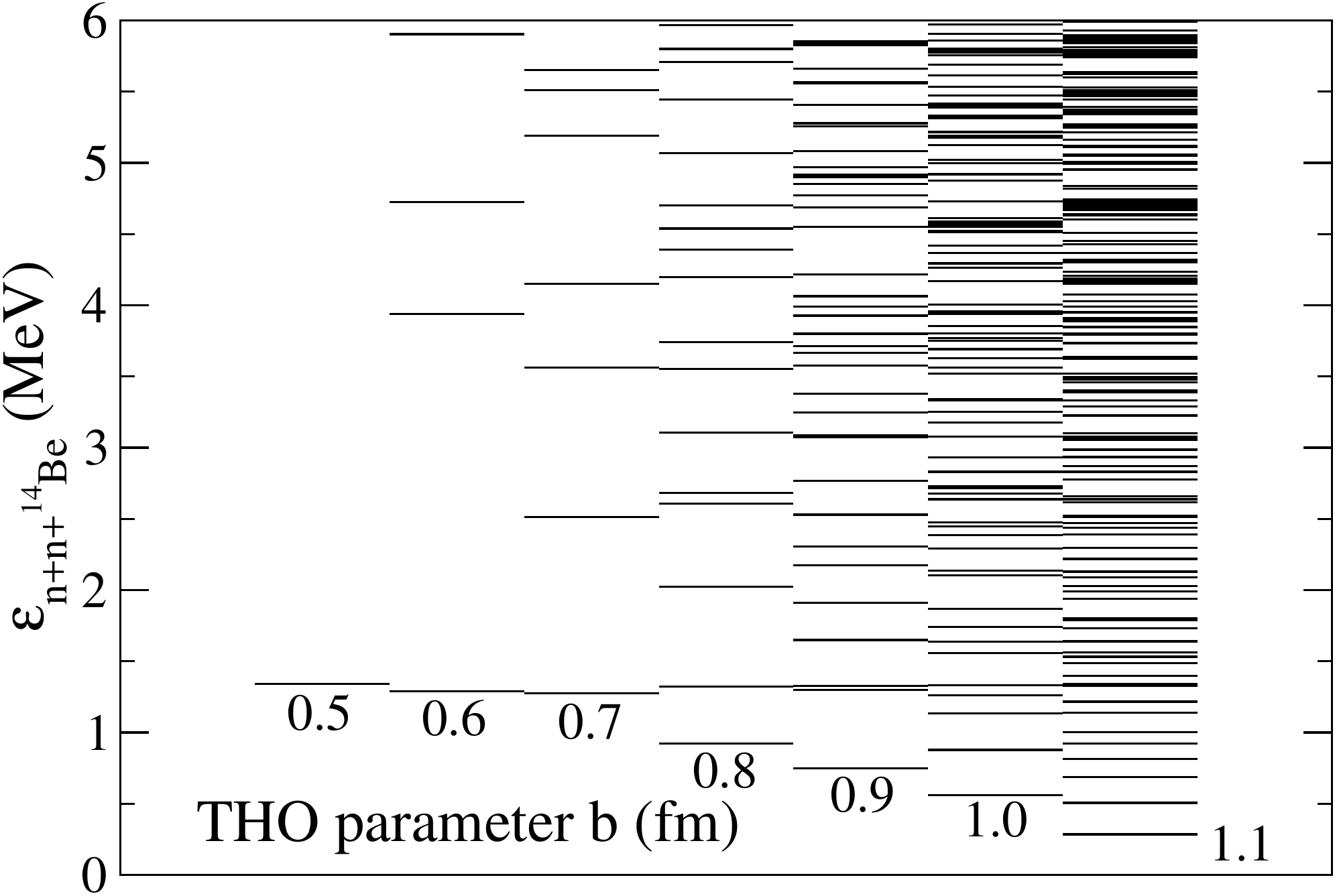} 
	\caption{As Fig.~\ref{fig:stabilization}, but changing the parameter $b$ while keeping $\gamma=2$ fm$^{1/2}$.}
	\label{fig:stabilization2}
\end{figure}

In this work, the three-body continuum problem is solved approximately within the three-body PS method using the THO basis. The parameters of the analytical transformation defining the basis control the level density after diagonalization~\cite{JCasal13}. Following the stabilization method by Hazi and Taylor~\cite{HaziTaylor70}, stable eigenstates close to resonance energies provide a good approximation of the inner part of the exact scattering wave function. The stability can be checked by changing the parameters $\gamma$ and $b$ of the transformation~\cite{JALay12}. However, it is worth noting that not any combination of these parameters is suitable for the purpose of this work. Since the ratio $\gamma/b$ controls the level density, as discussed in Sec.~\ref{sec:formalism}, very small $\gamma$ or large $b$ values give rise to a high concentration of states at low energy, which would make difficult the identification of a resonance. On the other hand, large $\gamma$ or small $b$ values go to the other limit, where the radial extension of the basis function is too small to cover the necessary range to describe physical systems. This would require many more radial excitations (i.e., larger $i_{max}$ values) to achieve convergence. In this sense, a compromise must be adopted and, in this work, the values of the parameters used are chosen by trial and error. 
In order to locate the ground-state resonance, the $^{16}$Be spectra obtained within different THO bases are shown in Figs.~\ref{fig:stabilization} and~\ref{fig:stabilization2}. In this calculations, either $b$ or $\gamma$ is fixed,
and the three-body problem is solved using different values of the other parameter and 
truncating the basis expansion with $K_{max}=30$ and $i_{max}=15$. From Fig.~\ref{fig:stabilization}, where $b$ is fixed to 0.7 fm, it is clear that a state around 1.3 MeV shows a rather stable pattern and, for $\gamma$ values above 1.8 fm$^{1/2}$, is well isolated from the rest of discretized continuum states. With $\gamma=2$ fm$^{1/2}$, the state has a variational minimum and can be used to study the ground-state properties. A similar behavior is found in Fig.~\ref{fig:stabilization}, now changing the parameter $b$. This is a solid evidence that a resonance has been identified. The 
PS approximation to analyze resonance properties of three-body systems was previously reported, for instance, for the 2$^+$ resonance in $^6$He~\cite{MRoGa05,JCasal13} or the 5/2$^-$ resonance in $^9$Be~\cite{Descouvemont15}.

\begin{figure}
	\centering
	\includegraphics[width=0.91\linewidth]{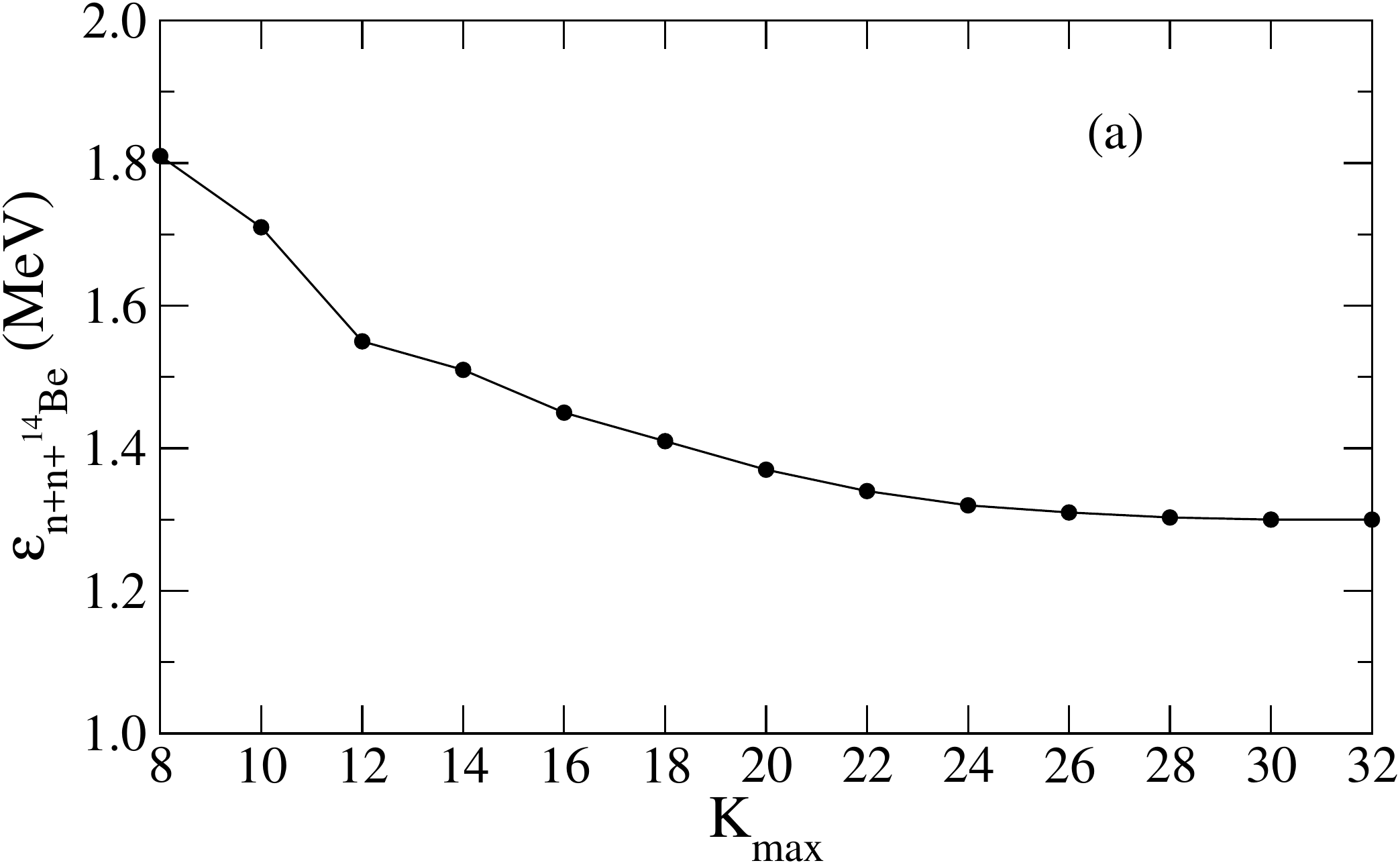} 
	
    \includegraphics[width=0.92\linewidth]{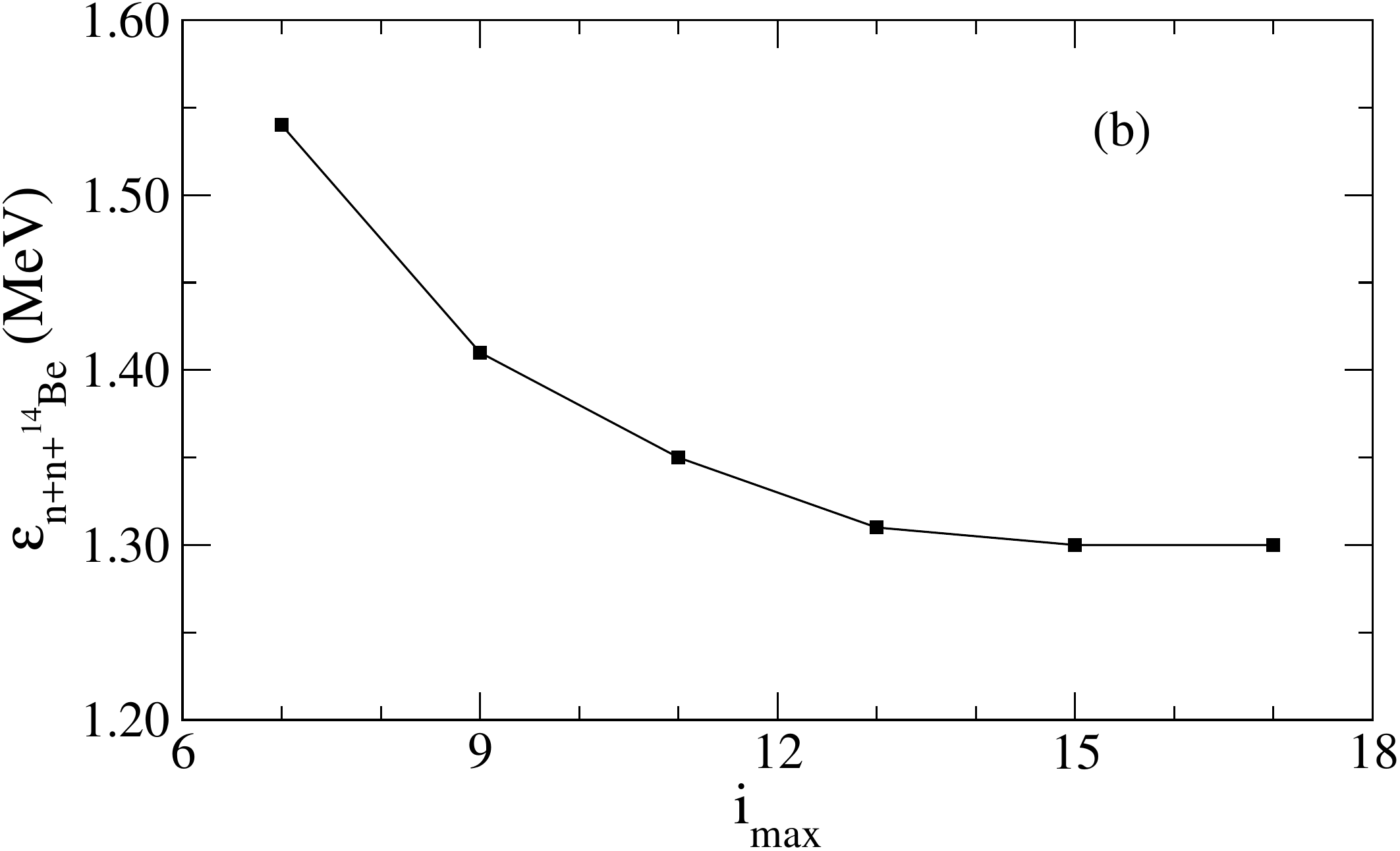} 
	\caption{Convergence of the $^{16}$Be ground-state energy as a function of (a) $K_{max}$ and (b) $i_{max}$. Calculations correspond to $\gamma=2.0$ fm$^{1/2}$.}
	\label{fig:conv16Be}
\end{figure}

The stability of the calculations is further clarified in Fig.~\ref{fig:conv16Be}, where the convergence of the ground-state energy as a function of the maximum hypermomentum $K_{max}$ and the number of hyperradial excitations $i_{max}$ is presented. This corresponds to the lowest PS obtained using $b=0.7$ fm and $\gamma=2$ fm$^{1/2}$, which is taken as an approximation of the resonance ground-state wave function. With $K_{max}=30$, corresponding to 136 $\beta$-channels in the wave function expansion~(\ref{eq:wf}), the resonance around 1.3 MeV is fully converged. It is also clear that $i_{max}=15$ hyperradial basis functions for each channel are sufficient to achieve convergence of the ground state. 

\begin{figure}
	\centering
	\includegraphics[width=1\linewidth]{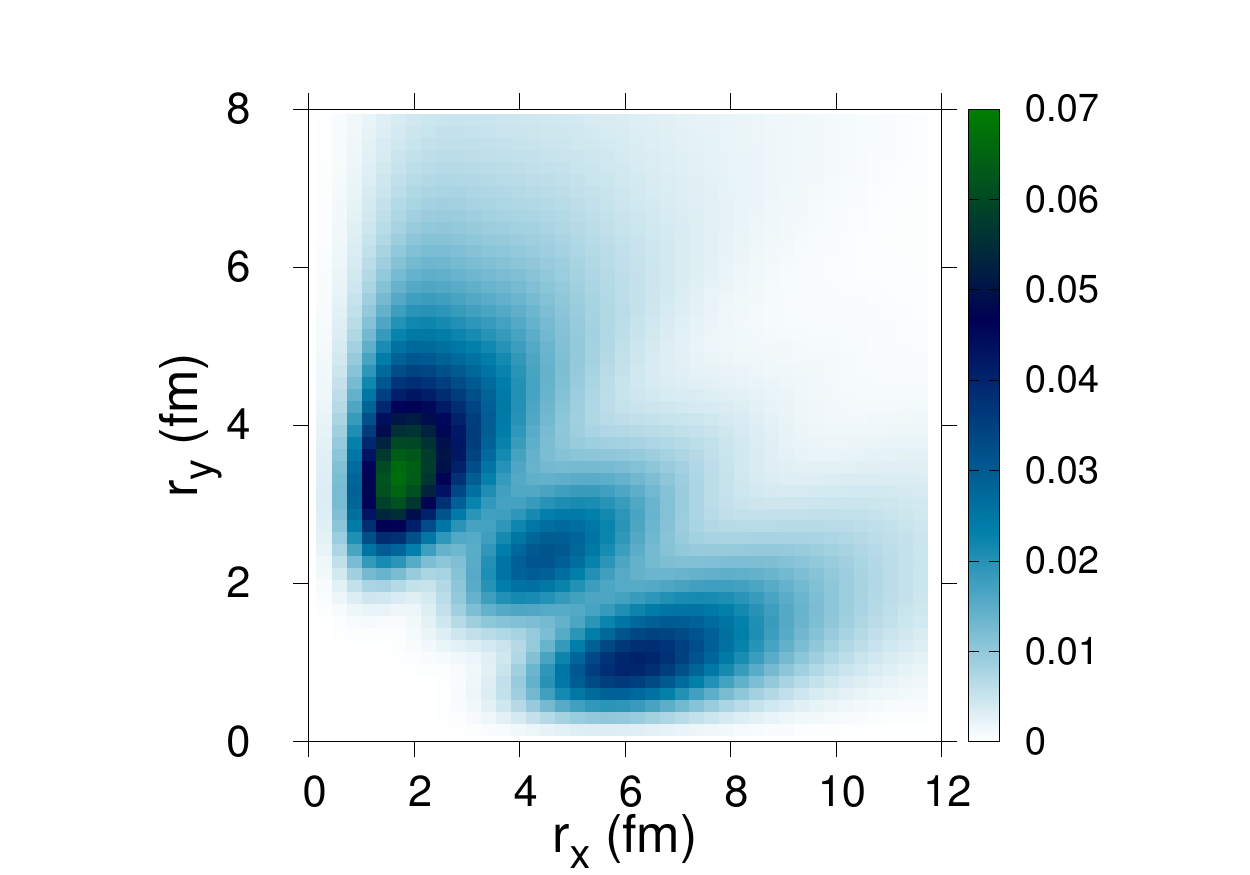} 
	\caption{(Color online) Ground-state probability of $^{16}$Be, with the scale given in fm$^{-2}$, as a function of $r_x\equiv r_{n\text{-}n}$ and $r_y\equiv r_{(nn)\text{-}^{14}{\rm Be}}$. The maximum corresponds to the dineutron configuration.}
	\label{fig:prob16Be}
\end{figure}

The spatial distribution of the valence neutrons in $^{16}$Be can be studied from the probability function defined by Eq.~(\ref{eq:Pxy}). Note that, with this definition, the two-dimensional contour plots avoid the axes, as opposed to the results presented in Ref.~\cite{lovell17} where the probability does not include the Jacobian. The ground-state probability is shown in Fig.~\ref{fig:prob16Be} as a function of $r_x\equiv r_{n\text{-}n}$ and $r_y\equiv r_{(nn)\text{-}^{14}{\rm Be}}$. The maximum at $r_x\simeq 2$ fm and $r_y\simeq 3.5$ fm corresponds to the dineutron configuration, while the other smaller peaks are typically associated with the triangle and cigar-like arrangements. From Fig.~\ref{fig:prob16Be}, it is clear that the dineutron component dominates the ground state of $^{16}$Be. A similar behavior was previously reported for the two-proton configuration in $^{17}$Ne~\cite{JCasal16}. This state is governed by relative $l_x=0$ components between the valence neutrons, which amount for 75\% of the total norm. From this results, it is possible to perform a transformation to the Jacobi set in which the $^{14}$Be core and a neutron are related by the $x$ coordinate. This transformation is related to the Raynal-Revai coefficients~\cite{RR70,IJThompson04}. In this set, the $n$-$^{14}$Be $d_{5/2}$ partial wave content of the ground state is 81\%. The total $d$-, $s$- and $p$-wave probabilities in this scheme are 85\%, 10\% and 4\%, respectively. The present calculations confirm the strong dineutron configuration in the $^{16}$Be ground state, which favors the picture of a correlated two-neutron emission. These findings agree with the experimental interpretation in Ref.~\cite{spyrou12} of $^{16}$Be as a ground-state dineutron emitter and are also consistent with the previous theoretical work~\cite{lovell17}. 

Note that, in Ref.~\cite{lovell17}, the actual $^{14}\text{Be}+n+n$ continuum was obtained within the $R$-matrix approach. In the present work, this problem has been approximated by solving a simple eigenvalue problem, which provides discrete PSs as a representation of the continuum. Results using the same three-body Hamiltonian are fully consistent, which supports the reliability of the PS method to study the properties of unbound three-body states. This can be achieved due to the versatility of the THO basis, whose analytical parameters enable the identification and analysis of single resonances. Note that, from the computational point of view, using the PS method is much less demanding than solving the actual continuum problem, such as in $R$-matrix calculations. Moreover, the present approach is general and can be easily applied to systems comprising any number of charged particles, for which the exact computation of scattering states is a well-known open problem. An application in this line will be presented in the following subsection for the case of $^6$Be.

\begin{figure}[t]
	\centering
	\includegraphics[width=1\linewidth]{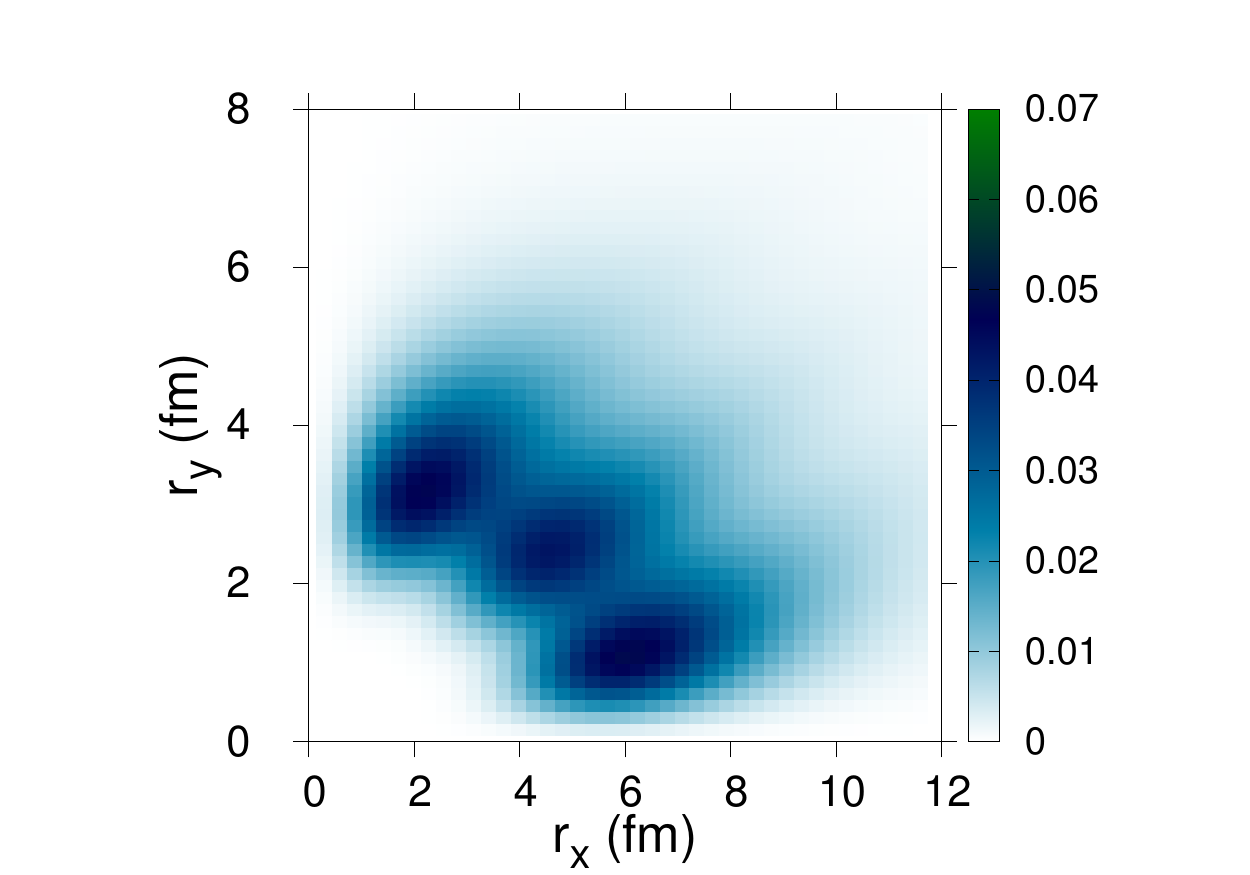} 
	\caption{(Color online) The same as Fig.~\ref{fig:prob16Be} but without $n$-$n$ interaction. The dineutron configuration no longer dominates.}
	\label{fig:prob16Be-noNN}
\end{figure}

While the properties of the binary subsystem $^{15}\text{Be}={^{14}\text{Be}}+n$ play a relevant role in shaping the properties of the compound system $^{16}$Be, the dominant dineutron configuration can be associated to the effect of the neutron-neutron interaction~\cite{lovell17}. This can be studied by diagonalizing the three-body Hamiltonian without the $NN$ potential. The ground-state probability corresponding to this unphysical solution is depicted in Fig.~\ref{fig:prob16Be-noNN}, where the same ${^{14}\text{Be}}+n$ interaction is employed. In this calculation, the three-body force has been adjusted to recover the same two-neutron separation energy. The fundamental difference with respect to the physical ground state in Fig.~\ref{fig:prob16Be} is the absence of a dominant dineutron configuration. Here, the dineutron and cigar-like contributions carry almost the same strength. In this case, the $n$-$n$ relative $l=0$ components are reduced to $\sim$50\%, while these valence neutrons occupy an almost pure $d_{5/2}$ orbit with respect to the core, i.e., $\sim$98\%. This illustrates that the strong dineutron character of the $^{16}$Be ground state is driven by the $NN$ interaction, and it is again consistent with the conclusions drawn in Ref.~\cite{lovell17}.

This work is focused on the identification of resonances and the study of their spatial correlations using a pseudostate approach. For the purpose of comparing with typical observables measured in the context of two-neutron emission~\cite{spyrou12}, it would be interesting to study: i) the relative energy distribution between the neutrons, and ii) the angular correlation between them. The later can be obtained through a change of coordinates, while the former could be achieved by performing the Fourier transform of the radial wave functions. Work along these lines is ongoing. Another interesting question rises regarding the determination of the width. In Ref.~\cite{lovell17}, the width of the $^{16}$Be ground state is obtained from the derivative of the scattering eigenphases. It has been shown (e.g., in Ref.~\cite{bartlett08}), that the width can also be estimated from the internal norm of the scattering wave functions. These energy distributions, however, are not available within the present pseudostate approach, so an estimation of the width in this context requires additional considerations. For two-body systems, integral formulae to compute the phase shifts and width of a state from a stabilized eigenstate are available~\cite{HaziTaylor70,TaylorHazi76}. Also in the three-body case, the treatment of resonances as ``quasibound'' states provides an estimation of the width from the internal part of the wave functions, as discussed in Ref.~\cite{Danilin98}. In that work, it is found that the quasibound wave function for a narrow resonance, obtained from actual continuum calculations, coincides with that obtained as a bound state to within 0.5\% of the total norm. This would open the possibility of studying the width by using pseudostates. However, this has yet to be explored in the context of the present THO calculations.

\subsection{$2p$ configuration in $^{6}$Be}
\label{sec:application2}
The three-body Hamiltonian for the $^4\text{He}+p+p$ system includes now, in addition to the binary nuclear potentials, the pair-wise Coulomb repulsion between the three interacting particles. Nevertheless, from the point of view of the PS method, the $^6$Be case is totally analogous to the one of $^{16}$Be discussed in the preceding subsection. Moreover, its mirror partner $^6$He can be described using the same method by just switching off the Coulomb part of the binary interactions. This enables a comparative study between both systems in a three-body scheme. 

\begin{figure}
	\centering
	\includegraphics[width=0.95\linewidth]{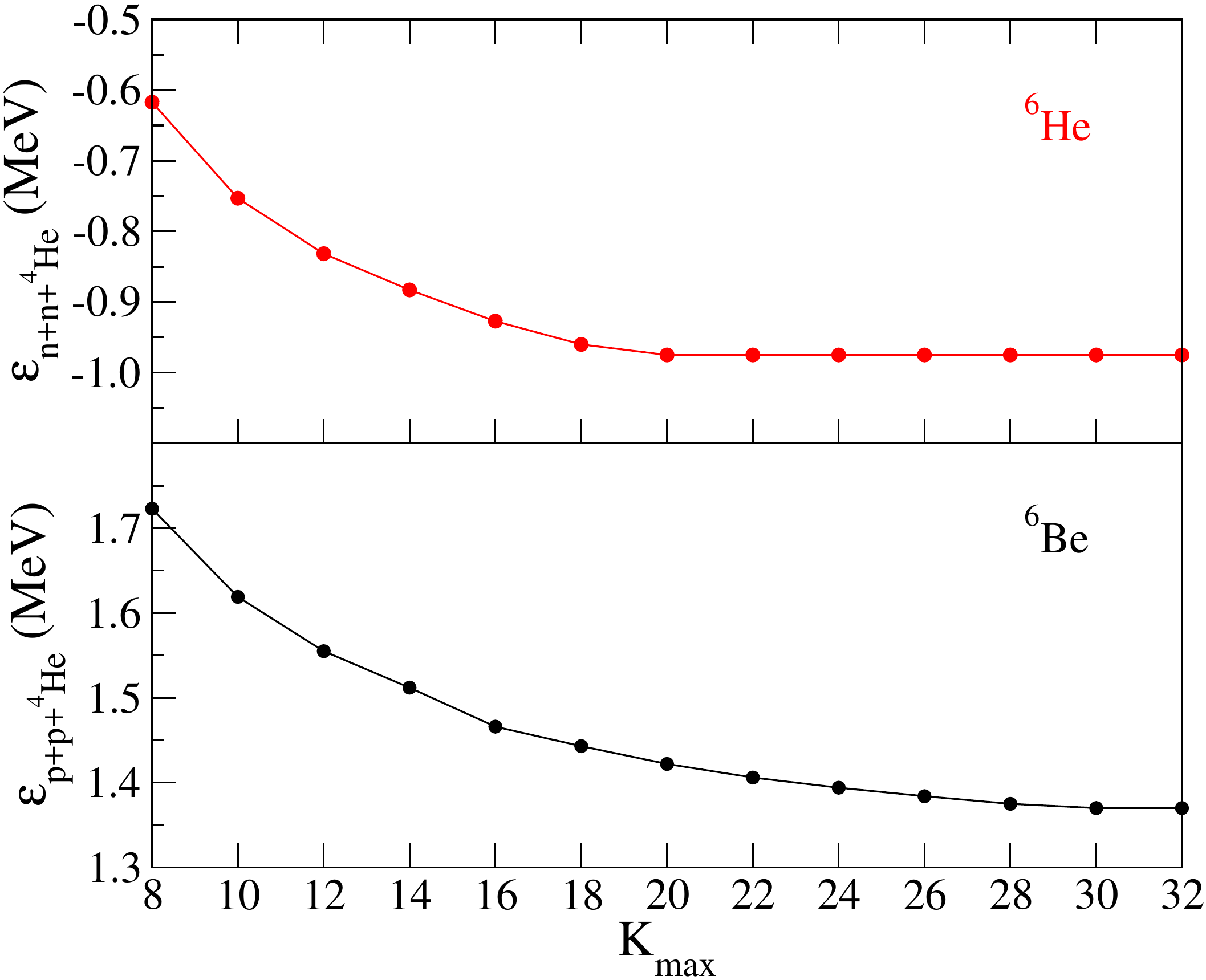} 

	\caption{Convergence of the $^{6}$He ground-state (upper panel) and $^{6}$Be (lower panel) as a function of the maximum hypermomentum $K_{max}$.}
	\label{fig:conv6He6Be}
\end{figure}

The $0^+$ ground states of $^6$He and $^6$Be are generated in a THO basis with the same parameters used for $^{16}$Be: $b=0.7$ fm and $\gamma=2$ fm$^{1/2}$. As in the previous case, this choice gives a stable PS in the continuum carrying the resonant properties of the $^6$Be ground-state. The position of the states is again adjusted using the three-body force introduced in Eq.~(\ref{eq:3b}) with $\rho_{\rm 3b}=6$ fm. The depths to reproduce the experimental two-neutron separation energy in $^6$He, 0.975 MeV~\cite{Brodeur12}, as well as the energy of the unbound $^6$Be are $v_{\rm 3b}=-2.35$ and $-$2.5 MeV, respectively. In Fig.~\ref{fig:conv6He6Be}, the convergence of the ground-state energy for both systems is shown as a function of $K_{max}$, which determines the size of the model space. It is clear that, within the PS approximation, the convergence of the bound state in $^6$He is achieved much faster than that of unbound systems. As in the previous case, the basis is set to $i_{max}=15$ hyperradial excitations, which was found to be sufficient to provide converged results.

\begin{figure}
	\centering
	\includegraphics[width=\linewidth]{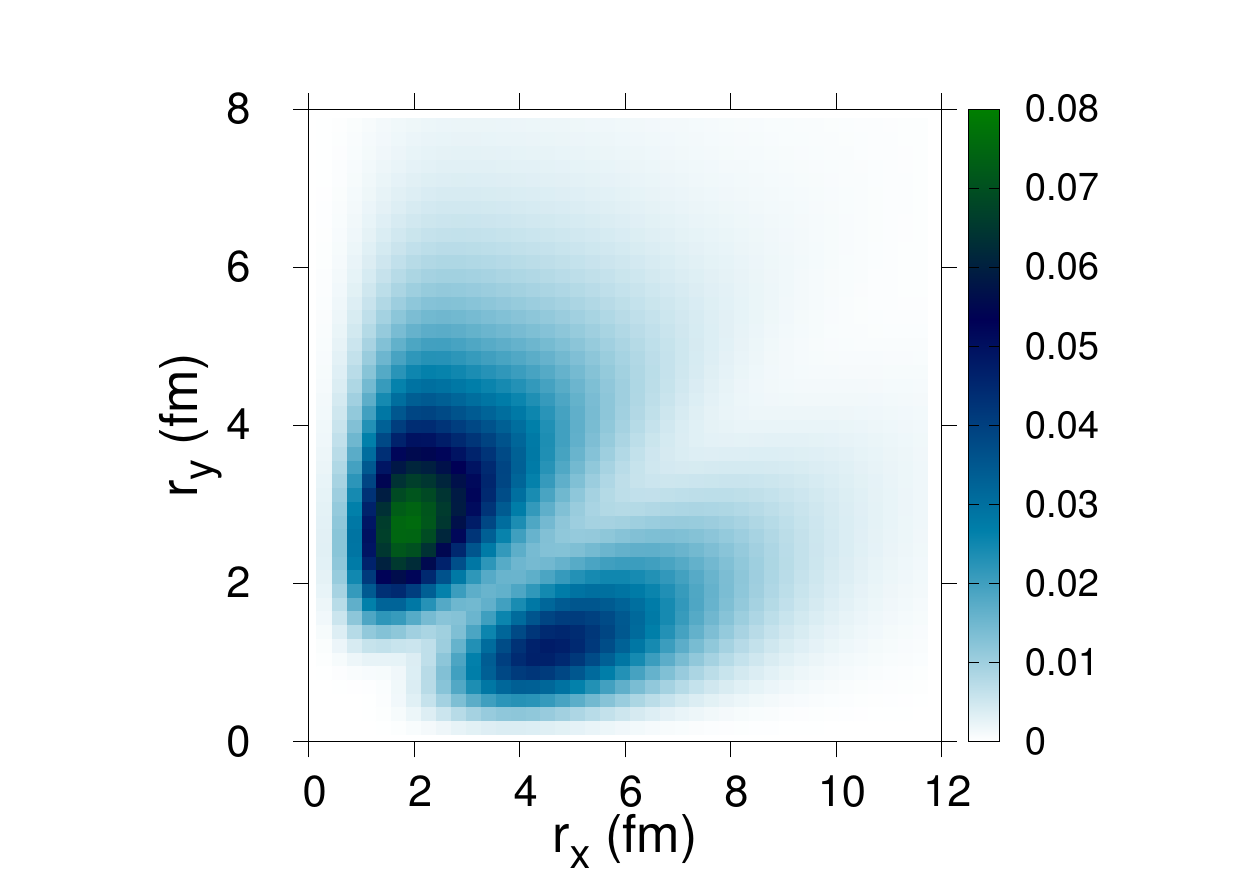} 
	\caption{(Color online) Ground-state probability of $^{6}$He  as a function of $r_x\equiv r_{n\text{-}n}$ and $r_y\equiv r_{(nn)\text{-}^{4}{\rm He}}$.}
	\label{fig:prob6He}
\end{figure}

\begin{figure}
	\centering
	\includegraphics[width=\linewidth]{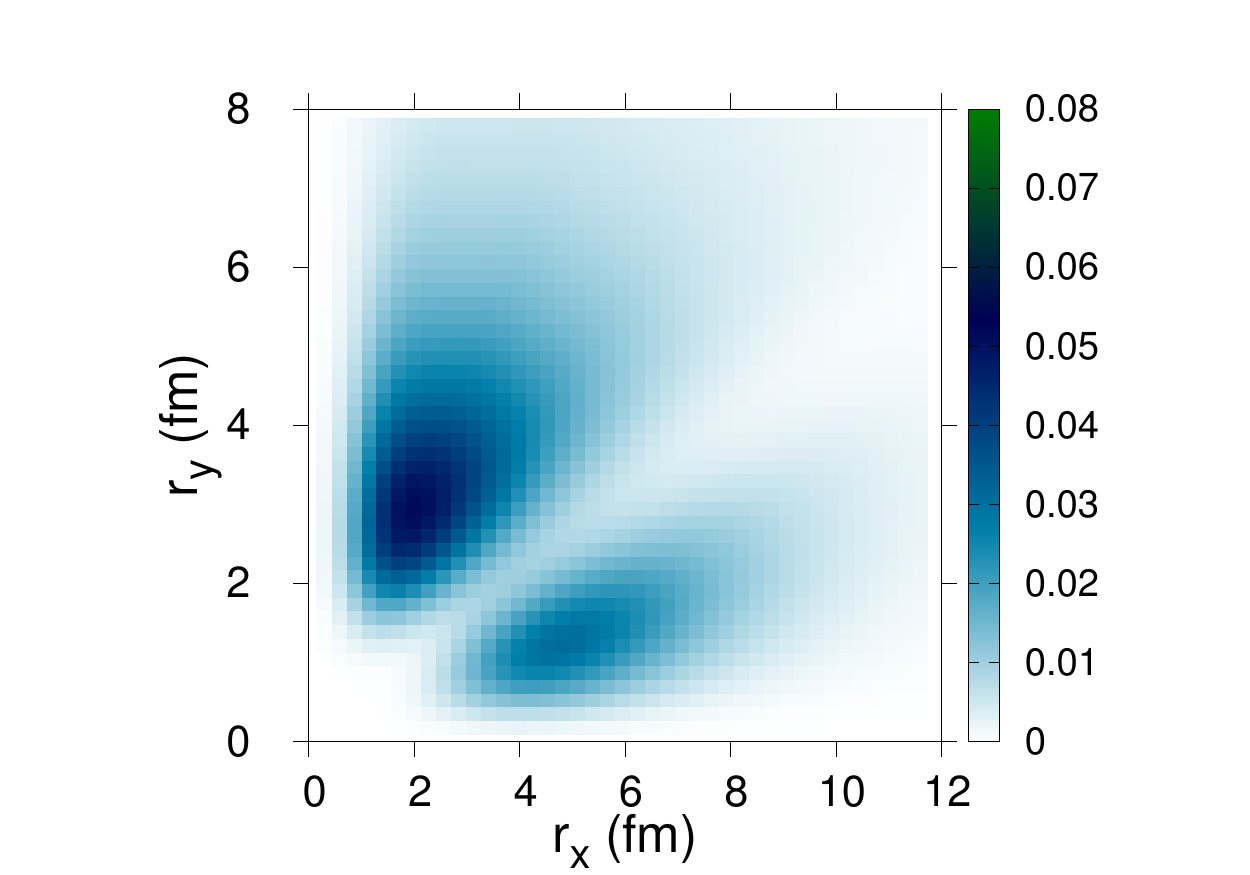} 
	\caption{(Color online) Ground-state probability of $^{6}$Be as a function of $r_x\equiv r_{p\text{-}p}$ and $r_y\equiv r_{(pp)\text{-}^{4}{\rm He}}$.}
	\label{fig:prob6Be}
\end{figure}

As in the previous example, the correlation between valence nucleons can be studied by plotting the ground-state probabilities. This is shown in Figs.~\ref{fig:prob6He} and~\ref{fig:prob6Be} for $^6$He and $^6$Be, respectively. The two-neutron halo in $^6$He presents the typical dineutron configuration~\cite{Zhukov93} around $r_x\simeq 2$ fm and $r_y\simeq 2.5$ fm. This is a clear signal of the strong correlations in the halo. The situation for $^6$Be is found to be analogous, with the absolute maximum corresponding to two protons close to each other at some distance apart from the $^4$He core.  The distribution is similar to the initial two-proton density presented in Ref.~\cite{oishi17}, where the $2p$ decay from $^6$Be is described as the time evolution of the valence protons in the spherical mean field generated by the core. The spatial distribution for $^6$Be is more diffuse than that for $^6$He, as it corresponds to an unbound system under the influence of the Coulomb repulsion between the three bodies. This is consistent with the results drawn in Ref.~\cite{SWang17} using the Gamov coupled-channel approach. The wave function contains 86\% (83\%) of relative $l_x=0$ components between the two valence protons (neutrons) in $^6$Be ($^6$He), and the nucleon-core $p_{3/2}$ content is close to 90\%. The present calculations confirm the strong diproton configuration in $^{6}$Be, in clear symmetry with the two-neutron halo of $^6$He. These results favor the picture of a correlated two-proton emission from the ground state of $^{6}$Be.

\section{Summary and conclusions}
\label{sec:conclusions}

Three-body calculations for the unbound $^{16}$Be ($^{14}\text{Be}+n+n$) and $^6$Be ($^4\text{He}+p+p$) systems have been carried out to study the correlations between the valence nucleons, in relation with two-nucleon radioactivity. Their ground states have been generated within the PS method using the analytical THO basis. This enables the identification of single resonances as discrete eigenstates in the continuum, which are stable with respect to the choice of the basis parameters. The models incorporate the GPT $NN$ interaction, realistic core-nucleon potentials adjusted to reproduce the known resonance energies of the binary subsystems $^{15}$Be($d_{5/2}$) and $^5$Li($p_{3/2}$), and also a phenomenological three-body force to adjust the position of the three-body states to the known experimental energies.

The ground-state probability distribution for $^{16}$Be presents a strong dineutron configuration, consistent with recent experimental observations. The present approach agrees with the conclusions from $R$-matrix calculations of actual scattering states. This supports the reliability of the PS method to study the ground-state properties of unbound three-body systems. The method is computationally less demanding and can be applied in general to systems comprising several charged particles. In this line, the ground state of $^6$Be shows a dominant diproton component, in clear symmetry with the two-neutron halo of its mirror partner $^6$He. For both $^{16}$Be and $^6$Be, the present results favor the picture of a correlated two-nucleon emission. 

From the present calculations, the next step involves the study of the energy correlation between valence nucleons, and the estimation of resonance widths within the PS approach. Other possible
applications of the 
method to study nucleon-nucleon correlations in unbound systems include the description of exotic oxygen isotopes such as $^{24}$O, $^{26}$O or $^{11}$O, the later being the mirror partner of the two-neutron halo $^{11}$Li. The decay from excited states of dripline nuclei, e.g., the resonances in $^6$He or $^{17}$Ne, and the influence of these correlations for reaction observables, could also be studied. Work along these lines is ongoing.

\begin{acknowledgments}
I am sincerely grateful to M.~Rodríguez-Gallardo and J.~M.~Arias for their constructive comments about the suitability of the present calculations. This work has been partially supported by the Spanish Ministerio de Economía y Competitividad under Projects No.~FIS2014-53448-C2-1-P and No.~FIS2014-51941-P, and by the European Union Horizon 2020 research and innovation program under Grant Agreement No.~654002.
\end{acknowledgments}

\bibliography{bibfile}

\end{document}